\documentclass[twocolumn,floats,floatfix,superscriptaddress,aps,pra]{revtex4}
\usepackage{amsfonts,amssymb,amsmath}
\usepackage{color,calc}
\usepackage[dvips]{graphicx}
\usepackage{bm}

\def\be{ \begin{equation} }
\def\ee{ \end{equation} }
\def\bea{ \begin{eqnarray} }
\def\eea{ \end{eqnarray} }
\def\bse{ \begin{subequations} }
\def\ese{ \end{subequations} }
\def\ba{ \begin{array} }
\def\ea{ \end{array} }

\def\i{\,\text{i}}
\def\e{\,\text{e}}
\def\i{i}
\def\e{e}

\def\to{\rightarrow}

\def\d{\text{d}}

\def\U{\mathbf{U}}

\def\A{\mathcal{A}}

\newcommand{\ket}[1]{\vert #1\rangle}
\def\i{{\rm{i}}}
\def\f{{\rm{f}}}
\def\phase{\phi}

\def\p{\mathcal{P}_{\theta}}
\def\eps{\epsilon}

\def\etal{\textit{et al.}}
\def\ibid{\textit{ibid.~}}

\begin{document}

\title{Arbitrarily accurate variable composite rotations of a quantum system}
\title{Arbitrarily accurate variable broadband composite rotations}
\title{Arbitrarily accurate variable broadband composite rotations on the Bloch sphere}
\title{Composite pulse sequences for arbitrarily accurate variable rotations on the Bloch sphere}
\title{Arbitrarily accurate variable rotations on the Bloch sphere by composite pulse sequences}

\author{Boyan T. Torosov}
\affiliation{Institute of Solid State Physics, Bulgarian Academy of Sciences, 72 Tsarigradsko chauss\'{e}e, 1784 Sofia, Bulgaria}

\author{Nikolay V. Vitanov}
\affiliation{Faculty of Physics, St Kliment Ohridski University of Sofia, James Bourchier 5 blvd, 1164 Sofia, Bulgaria}

\date{\today }

\begin{abstract}
Composite pulse sequences, which produce arbitrary pre-defined rotations of a two-state system at an angle $\theta$ on the Bloch sphere, are presented.
The composite sequences can contain arbitrarily many pulses and can compensate experimental errors in the pulse amplitude and duration to any desired order.
A special attention is devoted to two classes of $\pi/2$ sequences --- symmetric and asymmetric --- the phases of which are given by simple formulas in terms of rational multiples of $\pi$ for any number of constituent pulses.
This allows one to construct arbitrarily accurate $\pi/2$ composite rotations.
These $\pi/2$ composite sequences are used to construct three classes of arbitrarily long composite $\theta$ sequences by pairing two $\pi/2$ composite sequences, one of which is shifted by a phase $\pi-\theta$ with respect to the other one.
\end{abstract}

\maketitle


\section{Introduction\label{Sec:intro}}

Among the coherent control techniques in quantum physics, composite pulses have the unique advantage of combining
  ultrahigh accuracy similar to resonant techniques with robustness to parameter imperfections similar to adiabatic passage techniques.
Although the required total composite pulse area is typically a few times larger than the one used by the resonance techniques, it is still significantly less than the typical pulse areas in the adiabatic techniques.

Composite pulse sequences have been invented in nuclear magnetic resonance (NMR) nearly 50 years ago \cite{Levitt1979, Freeman1980, Levitt1982, Levitt1983, Tycko1984, Tycko1985, Shaka1984, Levitt1986, Wimperis1994}.
In the last decade they have enjoyed steadily growing interest in quantum information \cite{ions, Ivanov2011, Ivanov2015} and quantum optics \cite{Torosov2011PRA,Torosov2011PRL,Schraft2013, Genov2014PRL}.
In fact, the idea of composite sequences has been known in polarization optics since the 1940's.
By stacking several ordinary wave plates of the same or different material one can design either achromatic (broadband) polarization retarders or polarization filters,
 by rotating the plates at specific angles with respect to their fast polarization axes \cite{West1949,Destriau1949,Pancharatnam1955, Harris1964,McIntyre1968,Ivanov2012,Peters2012}.

The composite pulse sequence is a finite train of pulses with well-defined relative phases between them.
These phases are control parameters: they are determined by the desired excitation profile.
Composite pulses can shape the excitation profile in essentially any desired manner, an objective which is impossible with a single resonant pulse or adiabatic techniques.
In particular, one can create a broadband composite $\pi$ pulse, which delivers transition probability of 1 not only for a pulse area $\A=\pi$ and zero detuning $\Delta=0$, as a single resonant $\pi$ pulse, but also in some finite ranges around these values \cite{Freeman1980,Levitt1982,Levitt1983,Levitt1986,Wimperis1994,Torosov2011PRA,Torosov2011PRL}.
Thus a composite pulse can compensate the imperfections of a single real $\pi$ pulse and make it look like an ideal $\pi$ pulse.
Alternatively, narrowband composite pulses \cite{Tycko1984,Tycko1985,Shaka1984,Wimperis1994,Torosov2011PRA} squeeze the excitation profile around a certain point in the parameter space: they produce excitation that is more sensitive to parameter variations than a single pulse.
This enhanced sensitivity is suitable for applications in sensing, metrology and spatial localization in NMR spectroscopy.
A third family of composite pulses --- passband pulses --- combine the features of broadband and narrowband pulses: they provide highly accurate excitation inside a certain parameter range and negligibly small excitation outside it \cite{Kyoseva2013}.

Composite $\pi$ pulses produce a rotation by 180$^\circ$ on the Bloch sphere, or the NOT gate in quantum information language.
Of significant interest are also composite pulses, which produce partial rotation on the Bloch sphere at an angle $\theta$ --- composite $\theta$ pulses, or rotation gates in quantum information terms.
The most common of these are the $\pi/2$ composite pulses, which produce rotations at 90$^\circ$, but various composite pulses which produce robust rotations at arbitrary angle $\theta$ are also known in the literature.

In this paper, we present a set of composite pulse sequences, which produce a rotation at a pre-defined angle $\theta$, which is insensitive to variations in the amplitudes and the durations of the constituent pulses.
For $\theta=\pi/2$, which corresponds to the creation maximum coherence (ending at the equator) when starting from the south or north pole of the Bloch sphere, the phases of the constituent pulses are given by a very simple analytic formula, and so is the transition probability.
This allows one to explicitly assess the accuracy of these sequences.

Because for a single resonant pulse the transition probability is given by the well-known expression $p=\sin^2(\theta/2)$, where $\theta$ is the temporal pulse area,
 composite pulses that produce this transition probability are named $\theta$ pulses.
There are two classes of composite $\theta$ pulses, named variable and constant rotations \cite{Levitt1986}.
\emph{Variable}-rotation composite pulses compensate pulse area errors only in the transition probability $p$ (or the population inversion $w = 2p - 1 = -\cos\theta$).
\emph{Constant}-rotation composite pulses compensate pulse area errors in both the transition probability and the phases of the created superposition state (i.e., in the Bloch vector coherences $u$ and $v$).
The latter are obviously more demanding and require longer sequences for the same order of compensation.
The composite sequences derived and presented here belong to the class of variable rotations.
In Bloch sphere terms, they produce rotations of the Bloch vector from the south pole (inversion $w=-1$) to a pre-selected parallel; these rotations are very insensitive to pulse area fluctuations.
However, the exact position on the parallel is sensitive to such fluctuations.
Therefore, these $\theta$ pulses produce compensation of the polar rotation angle but do not compensate errors in the azimuthal angle.

This paper is organized as follows.
In Sec.~\ref{Sec:prime} we present prime composite $\theta$ pulses, which are directly derived from the overall composite propagator.
The important special case of composite $\pi/2$ sequences is considered in Sec.~\ref{Sec:half-pi} where a simple formula is presented for the composite phases for sequences of arbitrarily many pulses.
These composite $\pi/2$ sequences are used in Sec.~\ref{Sec:theta} to construct composite $\theta$ pulses --- termed twin composite pulses ---  by pairing two $\pi/2$ sequences, the second of which has a global phase shift to the first one.
In this manner, by using the arbitrarily long composite $\pi/2$ pulses we construct arbitrarily long composite $\theta$ pulses.
In Sec.~\ref{Sec:comparison} we compare the performance of our composite $\theta$ pulses to the most frequently used ones in the literature.
Finally, Sec.~\ref{Sec:conclusion} presents the conclusion.

\section{Prime composite $\theta$ pulses\label{Sec:prime}}

\subsection{General description}\label{sec:primes-general}

The derivation of the $\theta$ pulses is performed in a similar manner as in Refs.~\cite{Torosov2011PRA,Torosov2011PRL,Ivanov2011,Vitanov2011}; a brief description follows.
The propagator of a coherently driven two-state system is given by the SU(2) matrix,
\be\label{SU(2)}
\U_0 = \left[ \begin{array}{cc} a & b \\ -b^{\ast} & a^{\ast} \end{array}\right],
\ee
where $a$ and $b$ are  the (complex) Cayley-Klein parameters obeying $|a|^2+|b|^2=1$.
For exact resonance ($\Delta=0$), which we assume throughout, $a=\cos(\A/2) $, $b=-\i\sin(\A/2)$, where $\A$ is the temporal pulse area $\A=\int_{t_\i}^{t_\f}\Omega(t)\d t$.
For a system starting in state $\ket{1}$, the single-pulse transition probability is $p = |b|^2=\sin^2 (\A/2)$, and the population inversion reads $w=2p-1 = -\cos(\A)$.

A phase shift $\phase$ imposed on the driving field, $\Omega(t)\to\Omega(t)\e^{\i\phase}$, is imprinted onto the propagator as
\be\label{U phase}
\U_\phase = \left[ \begin{array}{cc} a & b \e^{\i\phase} \\ -b^{\ast}\e^{-\i\phase} & a^{\ast} \end{array}\right].
\ee
A train of $N$ pulses, each with area $\A_k$ and phase $\phase_k$,
\be
(\A_1)_{\phi_1} (\A_2)_{\phi_2} (\A_3)_{\phi_3} \cdots (\A_N)_{\phi_{N}},
\ee
 produces the propagator
\be\label{U^N}
\U^{(N)} = \U_{\phase_{N}}(\A_N) \cdots \U_{\phase_{3}}(\A_3) \U_{\phase_{2}}(\A_2) \U_{\phase_{1}}(\A_1).
\ee
The number of pulses $N$ can be odd or even.

\begin{figure}[t]
\includegraphics[width=0.80\columnwidth]{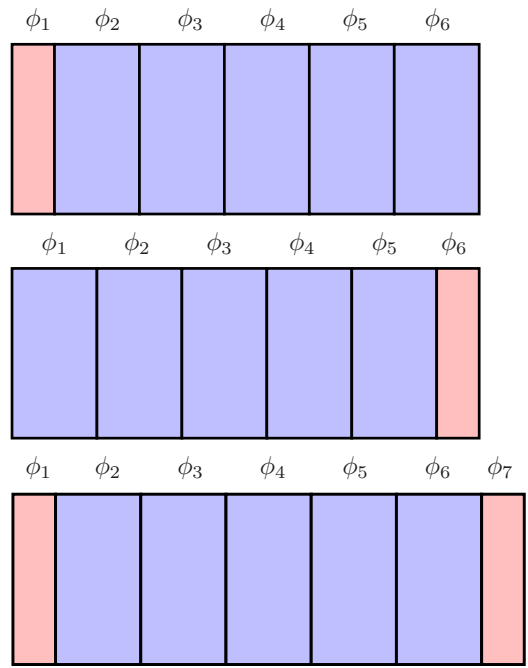}
\caption{
Composite pulse sequences considered in this paper, Eqs.~\eqref{3 sequences}.
Top: a single nominal $\pi/2$ pulse followed by a sequence of nominal $\pi$ pulses.
Middle: a sequence of nominal $\pi$ pulses followed by a single nominal $\pi/2$ pulse.
Bottom: a sequence of nominal $\pi$ pulses preceded and followed by single nominal $\pi/2$ pulses.
}
\label{fig:half-sequences}
\end{figure}

In this paper, based on numerical evidence, three types of composite pulse sequences are considered,
\bse\label{3 sequences}
\begin{align}
& A_{\phi_1} B_{\phi_2} B_{\phi_3} \cdots B_{\phi_{N-1}} B_{\phi_{N}}, \label{ABBB} \\
& B_{\phi_1} B_{\phi_2} B_{\phi_3} \cdots B_{\phi_{N-1}} A_{\phi_{N}}, \label{BBBA} \\
& A_{\phi_1} B_{\phi_2} B_{\phi_3} \cdots B_{\phi_{N-1}} A_{\phi_{N}}, \label{ABBA}
\end{align}
\ese
where $A=\pi(1+\eps)/2$ is a nominal (for zero error, $\eps=0$) $\pi/2$ pulse and $B=2A = \pi(1+\eps)$ is a nominal $\pi$ pulse.
The first two sequences are asymmetric with respect to the areas of the individual pulses, and the last one is symmetric.
The three types of sequences are illustrated in Fig.~\ref{fig:half-sequences}.

The main focus in this paper is at the transition probability $P^{(N)}  = |\U^{(N)}_{12}|^2$, or the population inversion $w^{(N)}  = 2P^{(N)}  - 1$.
The objective is to produce the pre-defined probability
\be
\p = \sin^2(\theta/2),
\ee
i.e., we wish to generate a composite $\theta$ pulse.

A single $\theta$ pulse gives the transition probability
\be\label{p-single}
P^{(1)} = \sin^2[\tfrac12 \theta (1+\eps)]  = \p [1 + \theta \cot(\theta/2) \eps + \cdots].
\ee
It is accurate only to first order in the error, $O(\eps)$.
The objective of the composite sequences is to make the composite probability $P^{(N)}$ equal to the target probability $\p$ at the nominal pulse area (i.e. for zero error) of the composite sequence, and insensitive to variations of the pulse area, i.e. insensitive to deviations in the highest possible order $m$ of the dimensionless error $\eps$, $O(\eps^m)$.

The procedure for derivation of the composite sequences is very simple.
First, the propagator \eqref{U^N} for the chosen composite sequence [among the ones of Eqs.~\eqref{3 sequences}] is calculated.
Then the transition probability $P^{(N)} = |\U^{(N)}_{12}|^2$ is computed and expanded in power series of $\eps$.
The coefficients in this power series depend on the composite phases $\phi_k$ ($k=1,2,\ldots,N$).
The composite phases are determined by setting the first, zeroth-order term $O(\eps^0)$ to the target probability $\p$, and annulling as many subsequent series coefficients as possible.
This amounts to solving a set of trigonometric equations.
The first nonzero term in the series beyond $O(\eps^0)$ (say, $\eps^m$) determines the error compensation order of the composite sequence, $O(\eps^m)$.

To this end, it is important to note that (i) if $\{\phi_1,\phi_2,\ldots \phi_N\}$ is a solution to the set of equations, then $\{-\phi_1,-\phi_2,\ldots -\phi_N\}$ is also a solution, which produces the same transition probability.
It is also obvious that (ii) if $\{\phi_1,\phi_2,\ldots \phi_N\}$ is a solution, then $\{\phi_1+2k_1\pi,\phi_2+2k_2\pi,\ldots \phi_N+2k_N\pi\}$ is also an equivalent solution, where $k_j$ are arbitrary integers.
Moreover, (iii) a composite sequence applied from left to right produces the same transition probability as the same sequence applied from right to left (although the two produce difference propagators, see Sec.~\ref{Sec:half-pi}).
Finally, (iv) if the same phase shift $\phi_0$ is added to all phases in the sequence, the transition probability remains the same (although the propagator changes).
These four features allow one, given a composite pulse sequence, to construct several other equivalent composite sequences, which deliver exactly the same transition probability.

The overall phase of the sequence is unimportant, hence one of the phases can be set to zero.
As such, in most cases we choose the phase of the first pulse, $\phi_1 = 0$.
Then all phases in such cases will be the relative phases with respect to the first pulse.


In the remainder of this section composite $\theta$ sequences of the symmetric type \eqref{ABBA} of up to six pulses are presented.
For sequences of two and three pulses the phases are derived analytically, while for more than three pulses they are derived numerically.
These composite pulses are termed ``prime'' because they are derived directly from the respective propagators by following the procedure outlined above.
Then, in Sec.~\ref{Sec:half-pi} analytic formulas for composite $\pi/2$ sequences of all three types \eqref{3 sequences} are obtained for an arbitrary number of pulses.
Finally, in Sec.~\ref{Sec:theta} these $\pi/2$ sequences are used to construct composite $\theta$ sequences of an arbitrary number of pulses.
The latter composite pulses are termed ``twin'' because they are composed of a pair of (prime) $\pi/2$ sequences.

\subsection{Two pulses}\label{sec:primes-2}

\begin{figure}[tb]
\includegraphics[width=0.95\columnwidth]{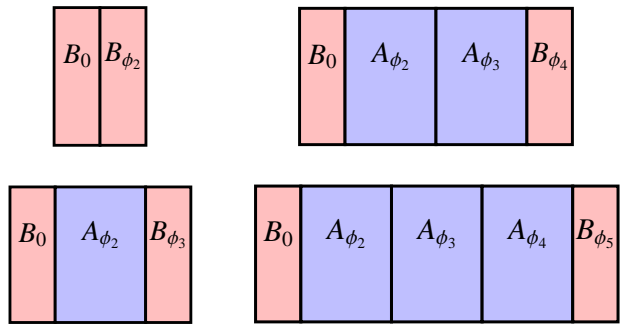}
\caption{
Prime composite $\theta$-pulses --- composite sequences of 2, 3, 4, and 5 pulses with the phases from Table \ref{Table:primes}.
}
\label{fig:sequences}
\end{figure}

The two-pulse composite sequence reads (cf. Fig.~\ref{fig:sequences})
\be\label{Theta2}
\Theta2 = A_{0} A_{\phi_2},
\ee
where, as before, $A=\pi(1+\eps)/2$ denotes a nominal $\pi/2$ pulse.
Hence the total nominal pulse area is just $\pi$.
The transition probability is readily calculated,
\be\label{P-2}
P^{(2)} = \cos ^2( \tfrac12\pi\eps ) \cos ^2 (\tfrac12 \phi_2).
\ee
Obviously, for a given target transition probability $\p$, the phase $\phi_2$ must be chosen as
\be
\phi_2 = \pi - \theta\quad \text{or} \quad \phi_2 = \pi + \theta,
\ee
where $\theta = \arccos(1-2\p)$.
For instance, transition probabilities of $\frac14$, $\frac12$ and $\frac34$ are realized by phases $\phi_2$ equal to $\frac23 \pi$, $\frac12 \pi$, and $\frac13 \pi$, respectively.

The composite transition probability \eqref{P-2} is accurate to the second order in $\eps$,
\be\label{P-2-eps}
P^{(2)} =  \p \left[1 - (\tfrac12\pi\eps)^2 + \cdots\right].
\ee
This is a quadratic improvement compared to a single $\theta$ pulse, which is accurate only to the first order, Eq.~\eqref{p-single}.

The reader will note that for $\p=1$, the composite phase is $\phi_2 = 0$ and Eq.~\eqref{P-2} reduces to the resonant $\pi$-pulse solution,
$P^{(2)} = \cos ^2( \tfrac12\pi\eps) = \sin^2( \tfrac12 \pi(1+\eps))$, as it should be the case.

Several two-pulse composite sequences of the type \eqref{Theta2} are listed in Table \ref{Table:primes}.

\begin{table*}
\begin{tabular}{|c|c|c|c|c|c|}
\hline
$\p$ & 2 pulses & 3 pulses  & 4 pulses & 5 pulses & 6 pulses \\
{} & $A_{0} A_{\phi_2}$ &
$A_{0} B_{\phi_2} A_{\phi_3}$ &
$A_{0} B_{\phi_2} B_{\phi_3} A_{\phi_4}$ &
$A_{0} B_{\phi_2} B_{\phi_3} B_{\phi_4} A_{\phi_5}$ &
$A_{0} B_{\phi_2} B_{\phi_3} B_{\phi_4} B_{\phi_5} A_{\phi_6}$ \\
\hline
{} & $\phi_1,\phi_2$ & $\phi_1,\phi_2,\phi_3$ & $\phi_1,\phi_2,\phi_3,\phi_4$ & $\phi_1,\phi_2,\phi_3,\phi_4,\phi_5$ & $\phi_1,\phi_2,\phi_3,\phi_4,\phi_5,\phi_6$  \\
\hline
$\frac1{10}$ & $0,0.7952 $ & $0,0.8204,1.4359 $  & $0,\frac23, 1.4618, 0.7952 $ & $0, 0.5033,1.6110,1.1032,1.7861 $ & $0, \frac25,\frac85,0.3952,1.1952,0.7952$ \\
$\frac18$ & $0,0.7699 $ & $0,0.8127 ,1.3954 $  & $0,\frac23 ,1.4366 ,0.7699 $ & $0,0.4891 ,1.5988 ,1.1258 ,1.8022 $ & $0, \frac25,\frac85,0.3699,1.1699,0.7699$ \\
$\frac16$ & $0,0.7323 $ & $0,0.8022 ,1.3367 $  & $0,\frac23 ,1.3990 ,0.7323 $ & $0,0.4698 ,1.5821 ,1.1599 ,1.8275 $ & $0, \frac25,\frac85,0.3323,1.1323,0.7323$ \\
$\frac15$ & $0,0.7048 $ & $0,0.7952 ,1.2952 $  & $0,\frac23 ,1.3715 ,0.7048 $ & ${0, 0.4569 , 1.5710 , 1.185 , 1.8467 }$ & $0, \frac25,\frac85,0.3048,1.1048,0.7048$ \\
$\frac14$ & $0,\frac23 $ & $0,0.7859 , 1.2386 $  & $0,\frac23 ,\frac43 ,\frac23 $ & $0, 0.4401 , 1.5564 , 1.2201 , 1.8743$ & $0, \frac25,\frac85,\frac{4}{15},\frac{16}{15},\frac23$ \\
$\frac13$ & $0,0.6082 $ & $0,0.7728 , 1.1537 $  & $0,\frac23 ,1.2748 ,0.6082$ & $0, 0.4162 , 1.5357 , 1.2743 , 1.9177 $ & $0, \frac25,\frac85,0.2082,1.0082,0.6082$ \\
$\frac12$ & $0,\frac12 $ & $0,\frac34 , 1$             & $0,\frac23 ,\frac76 ,\frac12 $ & $0, \frac38 , \frac32 , \frac{11}8 , 0$ & $0, \frac25,\frac85,\frac{1}{10},\frac{9}{10},\frac12 $ \\
$\frac23$ & $0,0.3918 $ & $0,0.7272 , 0.8463 $  & $0,\frac23 ,1.0585 ,0.3918 $ & $0, 0.3338 , 1.4643 , 1.4757 , 0.0823 $ & $0, \frac25,\frac85,1.9918,0.7918,0.3918$ \\
$\frac34$ & $0,\frac13 $ & $0,0.7141 , 0.7614 $  & $0,\frac23 , 1,\frac13 $           & $0, 0.3099 , 1.4436 , 1.5299 , 0.1257 $ & $0, \frac25,\frac85,\frac{29}{15},\frac{11}{15},\frac13 $ \\
$\frac45$ & $0,0.2952 $ & $0,0.7048 , 0.7048 $  & $0,\frac23 ,0.9618 ,0.2952 $ & $0, 0.2931 , 1.4291 , 1.565 , 0.1533 $ & $0, \frac25,\frac85,1.8952,0.6952,0.2952$ \\
$\frac56$ & $0,0.2677 $ & $0,0.6978 , 0.6633 $  & $0,\frac23 ,0.9344 ,0.2677 $ & $0, 0.2802 , 1.4179 , 1.5901 , 0.1725 $ & $0, \frac25,\frac85,1.8677,0.6677,0.2677$ \\
$\frac78$ & $0,0.2301 $ & $0,0.6873 , 0.6047 $  & $0,\frac23 ,0.8967 ,0.2301 $ & $0, 0.2609 , 1.4013 , 1.6242 , 0.1979 $ & $0, \frac25,\frac85,1.8301,0.6301,0.2301$ \\
$\frac9{10}$ & $0,0.2048 $ & $0,0.6796 , 0.5641 $  & $0,\frac23 , 0.8715, 0.2048 $ & $0, 0.2467, 1.3890, 1.6468, 0.2139$ & $0,\frac25,\frac85,1.8048,0.6048,0.2048$ \\
\hline
\end{tabular}
\caption{
Phases of composite pulse sequences which produce transition probability $\p=\sin^2(\theta/2)$.
All phases are given in units $\pi$.
}
\label{Table:primes}
\end{table*}

\subsection{Three pulses}\label{sec:primes-3}

The three-pulse composite sequences have the form (cf.~Fig.~\ref{fig:sequences})
\be\label{Theta3}
A_{\phi_1} B_{\phi_2}  A_{\phi_3}.
\ee
with $A=\pi(1+\eps)/2$ and $B=\pi(1+\eps)$.
Hence the total nominal pulse area is $2\pi$.

The expression for the transition probability $P^{(3)}$ is too cumbersome to be presented here.
The composite phases are determined by expanding $P^{(3)}$ vs the error $\eps$,
 setting the leading term to the desired probability $\p=\sin^2(\theta/2)$,
 and annulling the first nonzero term (which is of the order $O(\eps^2)$].
The resulting composite sequences read
\bse\label{phases-3}
\begin{align}
 & A_{\beta} B_{\alpha}  A_{-\beta}, \label{phases-3a}\\
 & A_{-\beta} B_{\alpha}  A_{\beta}, \label{phases-3b}\\
 & A_{0} B_{\alpha-\beta}  A_{-2\beta}, \label{phases-3c}\\
 & A_{0} B_{\alpha+\beta}  A_{2\beta}, \label{phases-3d}
\end{align}
\ese
with 
\bse
\begin{align}
\alpha &= \frac{\theta}{2},\\
\beta &= \cos ^{-1}\left(\sqrt{\p}-\sqrt{1-\p}\right) = \cos ^{-1}\left(\sqrt{2} \sin\frac{2\theta - \pi}{4}  \right).
\end{align}
\ese
The last three solutions can be obtained from the first one \eqref{phases-3a} by using the probability conserving transformations outlined in Sec.~\ref{sec:primes-general}.
The second solution \eqref{phases-3b} is obtained by reversing the order of the pulses in the first sequence \eqref{phases-3a}.
The last two solutions \eqref{phases-3c} and \eqref{phases-3d} are obtained from the first two solution \eqref{phases-3a} and \eqref{phases-3b} by subtracting the phase of the first pulse $\pm\beta$ from all phases.
Additional (equivalent) solutions can be generated by inverting the signs of all phases and using the other equivalence transformations from Sec.~\ref{sec:primes-general}.

All solutions \eqref{phases-3} are equivalent in the sense that they produce the same transition probability,
\be\label{P-3}
P^{(3)} = \p  \left(1-\sin^4 \frac{\pi  \eps}{2} \right) .
\ee
The composite transition probability \eqref{P-3} is accurate to the fourth order in $\eps$,
\be\label{P-3-eps}
P^{(3)} =  \p \left[1 - (\tfrac12{\pi\eps})^4 + \cdots\right].
\ee
Hence a quadratic improvement over the two-pulse sequence \eqref{Theta2}, which is accurate to $O(\eps^2)$, is obtained.

Several three-pulse composite sequences of the type \eqref{phases-3d} are listed in Table \ref{Table:primes}.


\subsection{Four pulses}\label{sec:primes-4}

We have derived two classes of four-pulse composite $\theta$ sequences.
The first class of sequences have the form (cf.~Fig.~\ref{fig:sequences})
\be\label{Theta4}
A_{\phi_1} B_{\phi_2} B_{\phi_3} A_{\phi_4},
\ee
with $A=\pi(1+\eps)/2$ and $B=\pi(1+\eps)$.
Hence the total nominal pulse area is $3\pi$.
%
The composite phases are determined as for the sequences of two and three pulses above.
Two exemplary composite sequences read
\bse\label{phases-4}
\begin{align}
 & A_0 B_{\frac23\pi} B_{\theta-\frac13\pi} A_{\theta+\pi}, \label{phases-4a}\\
 & A_0 B_{\frac23\pi} B_{\frac53\pi-\theta} A_{\pi-\theta}. \label{phases-4b}
\end{align}
\ese
Other equivalent solutions can be obtained from these two by using the probability conserving transformations outlined in Sec.~\ref{sec:primes-general}.
The solutions \eqref{phases-4} produce the same transition probability,
\be\label{P-4}
P^{(4)} = \p  \left(1-\sin^6 \frac{\pi  \eps}{2} \right) ,
\ee
which is accurate to the sixth order in $\eps$,
\be\label{P-4-eps}
P^{(4)} =  \p \left[1 - (\tfrac12{\pi\eps})^6 + \cdots\right].
\ee
Hence a qubic improvement over the two-pulse sequence \eqref{Theta2}, which is accurate to order $O(\eps^2)$, is obtained.
Several four-pulse composite sequences of the type \eqref{phases-4b} are listed in Table \ref{Table:primes}.

The second class of sequences have the form
\be\label{Theta4-half}
A_{\phi_1} A_{\phi_2} A_{\phi_3} A_{\phi_4}.
\ee
Hence the total nominal pulse area is $2\pi$.
Two exemplary composite sequences read
\bse\label{phases-4-half}
\begin{align}
 & A_0 A_{\frac12\pi} A_{\theta-\frac12\pi} A_{\theta+\pi}, \label{phases-4a-half}\\
 & A_0 A_{\frac32\pi} A_{\theta+\frac12\pi} A_{\theta+\pi}. \label{phases-4b-half}
\end{align}
\ese
Again, other equivalent sequences can be obtained from here by using the transformations outlined in Sec.~\ref{sec:primes-general}.
The sequences \eqref{phases-4-half} produce the transition probability
\be\label{P-4-half}
P^{(3)} = \p  \left(1-\sin^4 \frac{\pi \eps}{2} \right) .
\ee
The composite transition probability \eqref{P-4-half} is accurate to the fourth order in $\eps$, $O(\eps^4)$, the same as the three-pulse sequences with a nominal total pulse area of $2\pi$.

\begin{figure}[tbph]
\includegraphics[width=0.75\columnwidth]{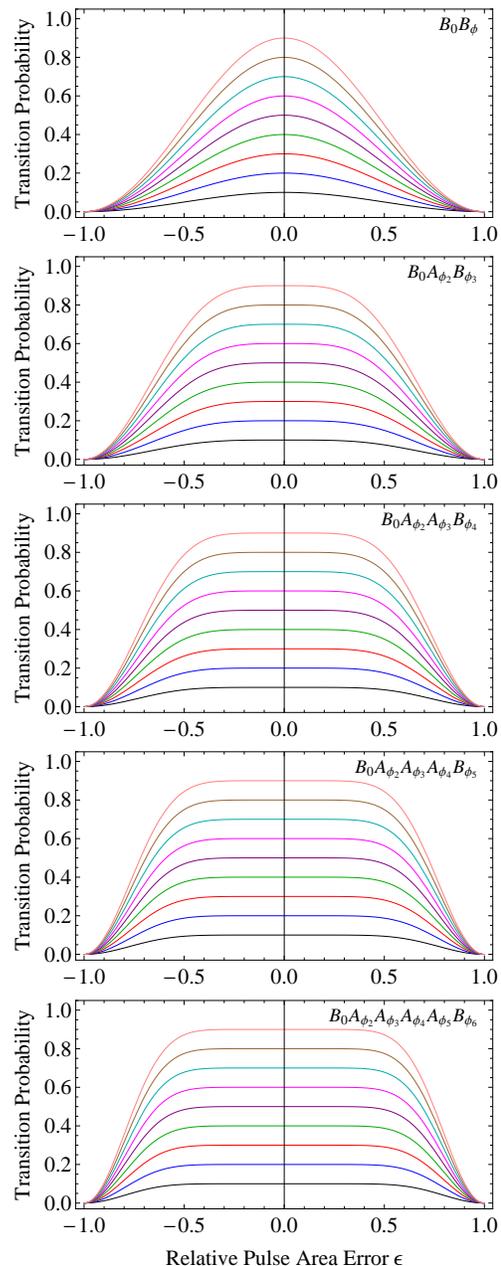}
\caption{
Transition probabilities induced by composite $\theta$-pulses --- composite sequences of 2 to 6 pulses with the phases from Table \ref{Table:primes}.
The curves in each frame show the transition probability locked at the levels $0.1, 0.2, \ldots, 0.9$.
}
\label{fig2}
\end{figure}

\subsection{More than four pulses}\label{sec:primes-more}

The five- and six-pulse composite sequences have the structure
\bse\label{Theta56}
\begin{align}
& A_{0} B_{\phi_2} B_{\phi_3} B_{\phi_4} A_{\phi_5}, \\
& A_{0} B_{\phi_2} B_{\phi_3} B_{\phi_4} B_{\phi_5} A_{\phi_6},
\end{align}
\ese
with $A=\pi(1+\eps)/2$ and $B=\pi(1+\eps)$.
Hence the total nominal pulse area is $4\pi$ and $5\pi$, respectively.
%
The composite phases $\phi_k$ are determined (numerically) by following the same procedure as above.
%
Several five- and six-pulse composite sequences are listed in Table \ref{Table:primes}.
There are multiple solutions for the composite phases leading to the same transition probability,
\bse
\begin{align}
P^{(5)} &= \p  \left(1-\sin^8 \frac{\pi  \eps}{2} \right), \\
P^{(6)} &= \p  \left(1-\sin^{10} \frac{\pi  \eps}{2} \right) .
\end{align}
\ese
They are accurate up to orders $O(\eps^8)$ and $O(\eps^{10})$, respectively.
Composite sequences of a larger number of pulses are derived similarly.
They are not presented here for the sake of brevity.

The transition profiles of the composite $\theta$ sequences of 2 to 6 pulses are presented in Fig.~\ref{fig2}.
Obviously, the larger the number of pulses, the flatter the excitation profile.


\section{Half-$\pi$ composite pulses\label{Sec:half-pi}}

As it is visible from Table \ref{Table:primes}, the phases for the $\pi/2$ composite pulses are given by rational multiples of $\pi$.
It turns out that there are simple analytic formulae for arbitrarily long $\pi/2$ composite pulse sequences, which are presented below.
There are two classes of such sequences --- symmetric and asymmetric (see Fig.~\ref{fig:half-sequences}) --- which are considered separately.

\subsection{Symmetric half-$\pi$ composite pulses}

The composite sequences of this type contain identical pulses of the same area $B = \pi (1+\eps)$ (nominal $\pi$-pulses) except for the first and last pulses, which have a half of this area, $A = \frac12\pi (1+\eps)$ (nominal $\pi/2$-pulses),
\be\label{half-pi-symmetric}
A_{\phi_1} B_{\phi_2}  B_{\phi_3} \cdots B_{\phi_{N-1}} A_{\phi_N},
\ee
see Fig.~\ref{fig:half-sequences}(bottom) and Fig.~\ref{fig:sequences}.
%
Obviously the total nominal pulse area is $(N-1)\pi$.
For arbitrary number of pulses $N$ (even or odd) the composite phases are
\be\label{phases-half-pi-symmetric}
\phi_k = \frac{(k-1)^2}{2(N-1)}\pi \quad (k=1,2, \ldots, N).
\ee
A few explicit examples of such $\pi/2$ composite pulses read
\bse\label{half-sym}
\begin{align}
&A_{0} A_{\frac12\pi},\label{half-sym-2}\\
&A_{0} B_{\frac14\pi} A_{\pi},\label{half-sym-3}\\
&A_{0} B_{\frac16\pi} B_{\frac23\pi} A_{\frac32\pi},\label{half-sym-4}\\
&A_{0} B_{\frac18\pi} B_{\frac12\pi} B_{\frac98\pi} A_{0},\label{half-sym-5}\\
&A_{0} B_{\frac1{10}\pi} B_{\frac25\pi} B_{\frac{9}{10}\pi} B_{\frac{8}{5}\pi} A_{\frac{1}{2}\pi}.\label{half-sym-6}
\end{align}
\ese
In the application of formula \eqref{phases-half-pi-symmetric} it is used that the phases are determined modulo $2\pi$.

The transition probability induced by these composite pulses reads
\be\label{half-probability-sym}
P^{(N)} = \tfrac12 - \tfrac12 \sin^{2N-2}\left(\tfrac12{\pi \eps}\right).
\ee
Obviously, these composite pulses are accurate to order $O(\eps^{2N-2})$,
\be\label{half-pN-sym}
P^{(N)} \sim \tfrac12 - \left(\tfrac12{\pi \eps}\right)^{2(N-1)} + O(\eps^{2N}) .
\ee

\begin{figure}[t]
\includegraphics[width=0.85\columnwidth]{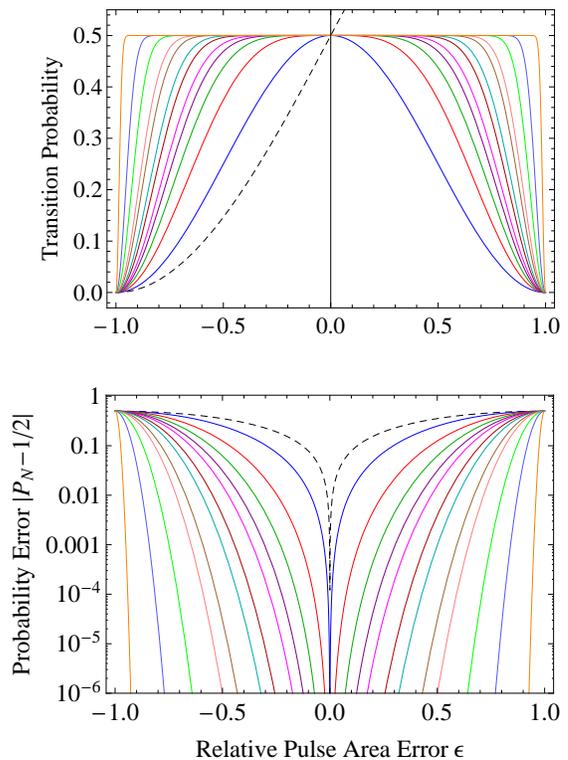}
\caption{
Transition probability for several symmetric $\pi/2$ composite sequences of the form \eqref{half-pi-symmetric} composed of 2, 3, 4, 5, 6, 8, 10, 15, 20, 40, 100 and 1000 pulses (from inside out, solid curves).
The composite phases are given by Eq.~\eqref{phases-half-pi-symmetric}.
The dashed curve shows the transition probability induced by a single pulse.
}
\label{fig-half-symmetric}
\end{figure}

Figure \ref{fig-half-symmetric} shows the transition probability for several symmetric $\pi/2$ composite sequences.
As the number of the pulses in the composite sequences increases the accuracy increases too [cf. Eq.~\eqref{half-pN-sym}], and the profiles become flatter and hence more robust to pulse area errors.

\subsection{Asymmetric half-$\pi$ composite pulses}

The composite sequences of this type contain identical nominal $\pi$-pulses, $B = \pi (1+\eps)$, except for the first pulse, which is a nominal $\pi/2$ pulse, $A = \frac12\pi (1+\eps)$,
\be\label{half-pi-asymmetric}
A_{\phi_1} B_{\phi_2}  B_{\phi_3} \cdots  B_{\phi_N},
\ee
see Fig.~\ref{fig:half-sequences}(top).
The total nominal pulse area is $(N-\frac12)\pi$.
The composite phases read
\be\label{phases-half-pi-asymmetric}
\phi_k = \frac{2(k-1)^2}{2N-1} \pi \quad (k=1,2,\ldots, N).
\ee
A few examples of such $\pi/2$ composite pulses read (using that the phases are determined modulo $2\pi$)
\bse\label{half-asym}
\begin{align}
&A_{0} B_{\frac23\pi},\label{half-asym-2}\\
&A_{0} B_{\frac25\pi} B_{\frac85\pi},\label{half-asym-3}\\
&A_{0} B_{\frac27\pi} B_{\frac87\pi} B_{\frac47\pi},\label{half-asym-4}\\
&A_{0} B_{\frac29\pi} B_{\frac89\pi} B_0 B_{\frac{14}9\pi},\label{half-asym-5}\\
&A_{0} B_{\frac2{11}\pi} B_{\frac8{11}\pi} B_{\frac{18}{11}\pi} B_{\frac{10}{11}\pi} B_{\frac{6}{11}\pi}.\label{half-asym-6}
\end{align}
\ese

\begin{figure}[t]
\includegraphics[width=0.85\columnwidth]{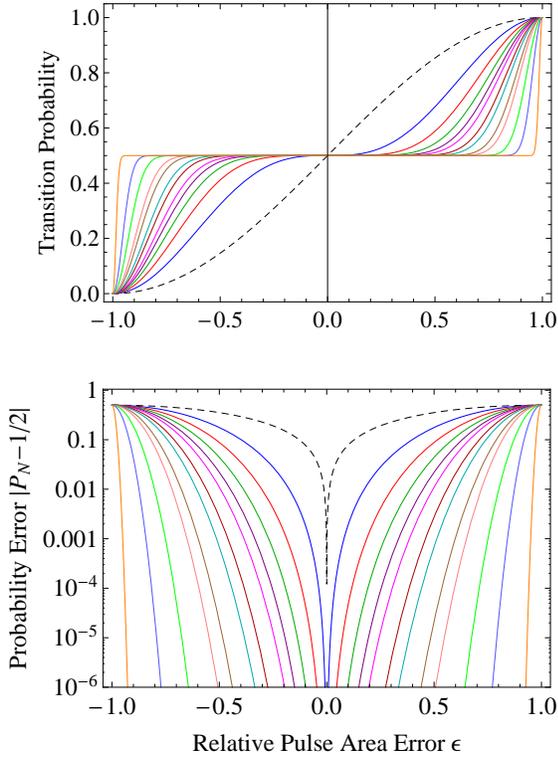}
\caption{
Transition probability for several asymmetric $\pi/2$ composite sequences of the form \eqref{half-pi-asymmetric} composed of 2, 3, 4, 5, 6, 8, 10, 15, 20, 40, 100 and 1000 pulses (from inside out, solid curves).
The composite phases are given by Eq.~\eqref{phases-half-pi-asymmetric}.
The dashed curve shows the transition probability induced by a single pulse.
}
\label{fig-half-asym}
\end{figure}

The transition probability has the simple form
\be\label{half-probability-asym}
P^{(N)} = \tfrac12 + \tfrac12 \sin^{2N-1}\left(\tfrac12\pi\eps\right) .
\ee
Obviously, it compensates errors up to order $O(\eps^{2N-1})$,
\be
P^{(N)} \sim \tfrac12 + \left(\tfrac12{\pi \eps}\right)^{2N-1} + O(\eps^{2N+1}).
\ee

Figure \ref{fig-half-asym} shows the transition probability for several asymmetric $\pi/2$ composite sequences.
As for the symmetric sequences, when the number of pulses in the composite sequences increases the profiles become flatter and hence more robust to pulse area errors.

\subsection{Comments}

The availability of analytic formulas for the composite phases for both the symmetric and asymmetric sequences allows one to write down immediately arbitrarily long $\pi/2$ composite sequences, which can be made accurate to any desired order.
In this sense, these sequences are \emph{arbitrarily accurate}.
Moreover, the extremely simple analytic formulae \footnote{The proof of Eqs. \eqref{half-probability-sym} and \eqref{half-probability-asym} follows the steps of the derivation in \cite{Torosov2018}.} for the transition probability \eqref{half-probability-sym} and \eqref{half-probability-asym} allow one to easily select a composite sequence, which compensates a given range of pulse area errors.
For example, transition probability in the range $(\frac12-10^{-4},\frac12+10^{-4})$ for admissible pulse area error $\eps$ of 0.1, 0.2 or 0.3 is delivered, respectively, by symmetric sequences of 3, 5, and 6 pulses, as well as asymmetric sequences of 4, 5, and 7 pulses.

Finally, it is important to note that if we have a composite sequence of the kind \eqref{half-pi-symmetric}, then the reversed sequence
\be\label{half-pi-sym-reversed}
A_{\phi_N} B_{\phi_{N-1}} \cdots B_{\phi_3} B_{\phi_2} A_{\phi_1},
\ee
will produce the same transition probability as the former one.
Likewise, the composite sequence \eqref{half-pi-asymmetric} and the reversed sequence [cf. Fig.~\ref{fig:half-sequences}(middle)]
\be\label{half-pi-asym-reversed}
 B_{\phi_N} B_{\phi_{N-1}} \cdots B_{\phi_3} B_{\phi_2} A_{\phi_1},
\ee
will produce the same transition probability.
Indeed, it can easily be shown that if the propagator for the sequence \eqref{half-pi-symmetric} is given by
\be\label{propagator}
\U = \left[\begin{array}{cc}
 a & b \\ -b^* & a^*
\end{array}\right],
\ee
then the propagator for the reversed sequence \eqref{half-pi-sym-reversed} reads
\be\label{propagatorR}
\U^R = \left[\begin{array}{cc}
 a^* & b \\ -b^* & a
\end{array}\right].
\ee
Since the off-diagonal elements of the two propagators are equal, the transition probabilities are equal too.
Similar conclusions hold for the mirror sequences \eqref{half-pi-asymmetric} and \eqref{half-pi-asym-reversed}.

\section{Arbitrarily accurate twin $\theta$ pulses\label{Sec:theta}}

\def\th{\vartheta}

Let us now consider a composite pulse sequence of the type \eqref{half-pi-symmetric} followed by its reversed counterpart \eqref{half-pi-sym-reversed}, with the phases of the latter shifted by the same phase $\th=\pi-\theta$ \footnote{A similar approach has been used by Levitt and Ernst \cite{Levitt1983} for construction of $\pi/2$ sequences.},
\be\label{half-pi-sym-reversed-theta}
A_{\phi_N+\th} B_{\phi_{N-1}+\th} \cdots B_{\phi_3+\th} B_{\phi_2+\th} A_{\phi_1+\th}.
\ee
The first sequence \eqref{half-pi-symmetric} will produce the propagator \eqref{propagator}, and the phase-shifted reversed sequence \eqref{half-pi-sym-reversed-theta} will produce the propagator \eqref{propagatorR}, but with phase shifts,
\be\label{propagatorR-shifted}
\U^R_{\th} = \left[\begin{array}{cc}
 a^* & b e^{i\th} \\ -b^* e^{-i\th} & a
\end{array}\right].
\ee
The propagator for the total twin sequence
\be\label{theta-sym-reversed}
A_{\phi_1} B_{\phi_2} \cdots B_{\phi_{N-1}} A_{\phi_N}
 A_{\phi_N+\th} B_{\phi_{N-1}+\th} \cdots B_{\phi_2+\th} A_{\phi_1+\th}
\ee
 reads
\be
\U^R_{\th} \U = \left[\begin{array}{cc}
 |a|^2 - |b|^2 e^{i \th } & a^* b (1+e^{i \th}) \\
 -a b^* (1+e^{-i \th}) & |a|^2 - |b|^2 e^{-i \th} \\
\end{array}\right] .
\ee
The transition probability is $|a b^* (1+e^{-i \th})|^2$, or
\be\label{prob-theta}
P = 4p (1-p) \cos^2(\tfrac12\th) = 4p (1-p) \sin^2(\tfrac12\theta),
\ee
where $p = |b|^2 = 1 - |a|^2$ is the transition probability for the single sequence described by the propagator \eqref{propagator}.
Therefore, if $p=\frac12$, then the transition probability is solely determined by the phase $\theta$,
\be
P = \cos^2(\tfrac12\th) = \sin^2(\tfrac12\theta).
\ee
Moreover, if the pulse generating the probability $p=\frac12$ is a composite-$\pi/2$ pulse, as in Eq.~\eqref{theta-sym-reversed}, then its properties will be imprinted onto this composite $\theta$ pulse.

Similar considerations apply for the asymmetric sequence \eqref{half-pi-asymmetric} followed by its phase-shifted reversed counterpart \eqref{half-pi-asym-reversed},
\be\label{theta-asym-reversed}
A_{\phi_1} B_{\phi_2} \cdots B_{\phi_{N-1}} B_{\phi_N}
 B_{\phi_N+\th} B_{\phi_{N-1}+\th} \cdots B_{\phi_2+\th} A_{\phi_1+\th}.
\ee
The transition probability for this twin composite sequence is given again by Eq.~\eqref{prob-theta} and it shares the properties of the composite sequences \eqref{half-pi-asymmetric} and \eqref{half-pi-asym-reversed}, which have formed it.

If the composite $\pi/2$ pulses are accurate to order $O(\eps^N)$, i.e., $P^{(N)} = \frac12 + c \eps^N$ then the composite $\theta$ pulse will be accurate to order $O(\eps^{2N})$ because $4P^{(N)}(1-P^{(N)}) = 4 (\frac12 +c\eps^N) (\frac12 -c\eps^N) = 1 - 4c^2 \eps^{2N}$.
Therefore, we can use the arbitrarily accurate composite $\pi/2$ pulses derived in the preceding section to derive arbitrarily accurate composite $\theta$-pulses by applying a composite $\pi/2$ pulse sequence $H$,
 followed by the same composite $\pi/2$ sequence $H^R_{\th}$ but applied in the reverse manner, with all its phases shifted by the same phase shift $\th$,
 \be\label{twin theta}
H_{0} H^R_{\th}.
 \ee
%
A few examples follow.

\begin{table*}
\begin{tabular}{|c|c|c|c|c|c|}
\hline
 $\theta=\frac14\pi$ ($\p=\frac{2-\sqrt{2}}{4}$) & $\frac13\pi$ ($\p=\frac14$) & $\frac23\pi$  ($\p=\frac34$) & $\theta=\frac34\pi$ ($\p=\frac{2+\sqrt{2}}{4}$) & \begin{tabular}{c} total \\ area \end{tabular} & \begin{tabular}{c} error \\ order \end{tabular} \\
\hline
     $A_{0}A_{\frac12\pi}A_{\frac54\pi}A_{\frac34\pi}$
 & $A_{0}A_{\frac12\pi}A_{\frac76\pi}A_{\frac23\pi}$
 & $A_{0}A_{\frac12\pi}A_{\frac56\pi}A_{\frac13\pi}$
 & $A_{0}A_{\frac12\pi}A_{\frac34\pi}A_{\frac14\pi}$
 & $2\pi$ & $O(\eps^4)$ \\
     $A_{0}B_{\frac23\pi}B_{\frac{17}{12}\pi}A_{\frac34\pi}$
 & $A_{0}B_{\frac23\pi}B_{\frac43\pi}A_{\frac23\pi}$
 & $A_{0}B_{\frac23\pi}B_{\pi}A_{\frac13\pi}$
 & $A_{0}B_{\frac23\pi}B_{\frac{11}{12}\pi}A_{\frac14\pi}$
 & $3\pi$ & $O(\eps^6)$ \\
     $B_{0}A_{\frac43\pi}A_{\frac1{12}\pi}B_{\frac34\pi}$
 & $B_{0}A_{\frac43\pi}A_{0}B_{\frac23\pi}$
 & $B_{0}A_{\frac43\pi}A_{\frac53\pi}B_{\frac13\pi}$
 & $B_{0}A_{\frac43\pi}A_{\frac{19}{12}\pi}B_{\frac14\pi}$
 & $3\pi$ & $O(\eps^6)$ \\
     $A_{0}B_{\frac14\pi}A_{\pi}A_{\frac74\pi}B_{\pi}A_{\frac34\pi}$
 & $A_{0}B_{\frac14\pi}A_{\pi}A_{\frac53\pi}B_{\frac{11}{12}\pi}A_{\frac23\pi}$
 & $A_{0}B_{\frac14\pi}A_{\pi}A_{\frac43\pi}B_{\frac{7}{12}\pi}A_{\frac13\pi}$
 & $A_{0}B_{\frac14\pi}A_{\pi}A_{\frac54\pi}B_{\frac12\pi}A_{\frac14\pi}$
 & $4\pi$ & $O(\eps^8)$ \\
     $A_{0}B_{\frac25\pi}B_{\frac85\pi}B_{\frac7{20}\pi}B_{\frac{23}{20}\pi}A_{\frac34\pi}$
 & $A_{0}B_{\frac25\pi}B_{\frac85\pi}B_{\frac4{15}\pi}B_{\frac{16}{15}\pi}A_{\frac23\pi}$
 & $A_{0}B_{\frac25\pi}B_{\frac85\pi}B_{\frac{29}{15}\pi}B_{\frac{11}{15}\pi}A_{\frac13\pi}$
 & $A_{0}B_{\frac25\pi}B_{\frac85\pi}B_{\frac{37}{20}\pi}B_{\frac{13}{20}\pi}A_{\frac14\pi}$
 & $5\pi$ & $O(\eps^{10})$ \\
     $B_{0}B_{\frac45\pi}A_{\frac25\pi} A_{\frac{23}{20}\pi}B_{\frac{31}{20}\pi}B_{\frac34\pi}$
 & $B_{0}B_{\frac45\pi}A_{\frac25\pi} A_{\frac{16}{15}\pi}B_{\frac{22}{15}\pi}B_{\frac23\pi}$
 & $B_{0}B_{\frac45\pi}A_{\frac25\pi} A_{\frac{11}{15}\pi}B_{\frac{17}{15}\pi}B_{\frac13\pi}$
 & $B_{0}B_{\frac45\pi}A_{\frac25\pi} A_{\frac{13}{20}\pi}B_{\frac{21}{20}\pi}B_{\frac14\pi}$
 & $5\pi$ & $O(\eps^{10})$ \\
\hline
\end{tabular}
\caption{
Several composite twin $\theta$ pulses of the type \eqref{twin theta}, which produce transition probability $\p=\sin^2(\frac12 \theta)$.
$A$ denotes a nominal $\pi/2$ pulse and $B$ a nominal $\pi$ pulse.
The sequences listed in each row are derived from Eqs.~\eqref{twin-sym-4}, \eqref{twin-asym-4}, \eqref{twin-asym-2-4}, \eqref{twin-sym-6}, \eqref{twin-asym-6}, and \eqref{twin-asym-2-6}, respectively.
}\label{table:twin}
\end{table*}

(i) By using the symmetric $\pi/2$ composite sequences \eqref{half-sym} we construct the twin $\theta$ sequences
\bse\label{twin-sym}
\begin{align}
& A_0 A_{\frac12\pi} A_{\frac32\pi-\theta} A_{\pi-\theta}, \label{twin-sym-4}\\
& A_0 B_{\frac14\pi} A_{\pi} A_{2\pi-\theta} B_{\frac54\pi-\theta} A_{\pi-\theta}, \label{twin-sym-6}\\
& A_0 B_{\frac16\pi} B_{\frac23\pi} A_{\frac32\pi} A_{\frac12\pi-\theta} B_{\frac53\pi-\theta} B_{\frac76\pi-\theta} A_{\pi-\theta},\\
& A_0 B_{\frac18\pi} B_{\frac12\pi} B_{\frac98\pi} A_{0} A_{\pi-\theta} B_{\frac18\pi-\theta} B_{\frac32\pi-\theta} B_{\frac98\pi-\theta} A_{\pi-\theta}.
\end{align}
\ese

(ii) Likewise, by using the asymmetric $\pi/2$ composite sequences \eqref{half-asym} we construct the twin $\theta$ sequences
\bse\label{twin-asym}
\begin{align}
& A_0 B_{\frac23\pi} B_{\frac53\pi-\theta} A_{\pi-\theta}, \label{twin-asym-4}\\
& A_0 B_{\frac25\pi} B_{\frac85\pi} B_{\frac35\pi-\theta} B_{\frac75\pi-\theta} A_{\pi-\theta}, \label{twin-asym-6}\\
& A_0 B_{\frac27\pi} B_{\frac87\pi} B_{\frac47\pi} B_{\frac{11}7\pi-\theta} B_{\frac17\pi-\theta} B_{\frac97\pi-\theta} A_{\pi-\theta},\\
& A_0 B_{\frac29\pi} B_{\frac89\pi} B_{0} B_{\frac{14}9\pi} B_{\frac{5}9\pi-\theta} B_{\pi-\theta} B_{\frac{17}9\pi-\theta} B_{\frac{11}9\pi-\theta} A_{\pi-\theta}.
\end{align}
\ese

(iii) The asymmetric $\pi/2$ composite sequences can be applied in the opposite order to case (ii), and hence we obtain other twin $\theta$ sequences
%
\bse\label{twin-asym-2}
\begin{align}
& B_{0}A_{\frac43\pi}  A_{\frac13\pi-\theta}B_{\pi-\theta}, \label{twin-asym-2-4}\\
& B_{0}B_{\frac45\pi}A_{\frac25\pi}  A_{\frac75\pi-\theta}B_{\frac95\pi-\theta}B_{\pi-\theta}, \label{twin-asym-2-6}\\
& B_{0}B_{\frac47\pi} B_{\frac{12}7\pi}A_{\frac{10}7\pi}   A_{\frac37\pi-\theta} B_{\frac57\pi-\theta} B_{\frac{11}7\pi-\theta} B_{\pi-\theta},\\
& B_{0}B_{\frac{4}9\pi}B_{\frac43\pi}B_{\frac23\pi}A_{\frac{4}9\pi}   A_{\frac{13}9\pi-\theta}B_{\frac53\pi-\theta}B_{\frac13\pi-\theta}B_{\frac{13}9\pi-\theta}B_{\pi-\theta}.
\end{align}
\ese

Table \ref{table:twin} shows a few examples of twin composite sequences.

(iv) By twinning two composite $\pi/2$ sequences with $\theta=\pi/2$, we can construct longer composite $\pi/2$ sequences; the shortest of these are [cf.~Eqs.~\eqref{twin-sym-4}, \eqref{twin-asym-4}, and \eqref{twin-asym-2-4}]
\bse\label{twin-half}
\begin{align}
& A_0 A_{\frac12\pi} A_{\pi} A_{\frac12\pi}, \label{twin-sym-half}\\
& A_0 B_{\frac23\pi} B_{\frac76\pi} A_{\frac12\pi}, \label{twin-asym-half}\\
& B_{0}A_{\frac43\pi}  A_{\frac{11}6\pi}B_{\frac12\pi}. \label{twin-asym-2-half}
\end{align}
\ese
They have the same error orders as the prime $\pi/2$ sequences in Sec.~\ref{Sec:half-pi} with the same nominal total pulse area.

(v) By twinning two composite $\pi/2$ sequences with $\theta=\pi$, we can construct composite $\pi$ sequences; the shortest of these are [cf.~Eqs.~\eqref{twin-sym-4}, \eqref{twin-asym-4}, and \eqref{twin-asym-2-4}]
\bse\label{twin-pi}
\begin{align}
& A_0 A_{\frac12\pi} A_{\frac12\pi} A_0, \label{twin-sym-pi}\\
& A_0 B_{\frac23\pi} B_{\frac23\pi} A_{0}, \label{twin-asym-pi}\\
& B_{0}A_{\frac43\pi}  A_{\frac43\pi}B_{0}. \label{twin-asym-2-pi}
\end{align}
\ese
These twin $\pi$ pulses have been discussed in an earlier publication \cite{Torosov2018}.

\section{Comparison with existing $\theta$ pulses\label{Sec:comparison}}

\subsection{Prime $\theta$ pulses}

There are several composite $\theta$ pulses in the literature, which produce variable or constant rotations \cite{Levitt1986}.

\subsubsection{Variable rotation pulses}

The first composite $\pi/2$ pulse has been introduced by Freeman \etal~\cite{Freeman1980}, which is a sequence of two nominal $\pi/2$ pulses, with a relative phase shift of $\pi/2$,
\be\label{Freeman1980}
A_{0} A_{\frac12\pi},
\ee
with $A= \pi (1+\eps)/2$.
This is the first member of the symmetric $\pi/2$ class, Eq.~\eqref{half-sym-2}.
It is accurate up to $O(\eps^2)$.

Another famous $\pi/2$ pulse is the two-pulse sequence of Levitt \cite{Levitt1982}, which contains a nominal $\pi/2$ pulse and a nominal $\pi$ pulse, with a relative phase shift of $\frac23\pi$,
\be
A_{0} B_{\frac23\pi},
\ee
with $A= \pi (1+\eps)/2$ and $B=\pi (1+\eps)$.
It is the first member of the present asymmetric $\pi/2$ class, Eq.~\eqref{half-asym-2}.
It is accurate up to  $O(\eps^3)$.

\subsubsection{Constant rotation pulses}

Wimperis \cite{Wimperis1994} introduced the (now very popular) BB1 composite $\theta$ pulse,
\be\label{WimpCP}
\Theta_{0} B_{\chi} B_{3\chi} B_{3\chi} B_{\chi},
\ee
with $B=\pi (1+\eps)$, $\Theta=\theta (1+\eps)$, and $\chi=\cos^{-1}(-\theta/(4\pi))$.
For $\theta = \pi/2$ we have $\chi=\cos^{-1}(-1/8)\approx 0.5399\pi \approx 97.18^{\circ}$.
The composite pulse \eqref{WimpCP} is accurate up to order $O(\eps^3)$.

\begin{figure}[tb]
	\includegraphics[width=0.95\columnwidth]{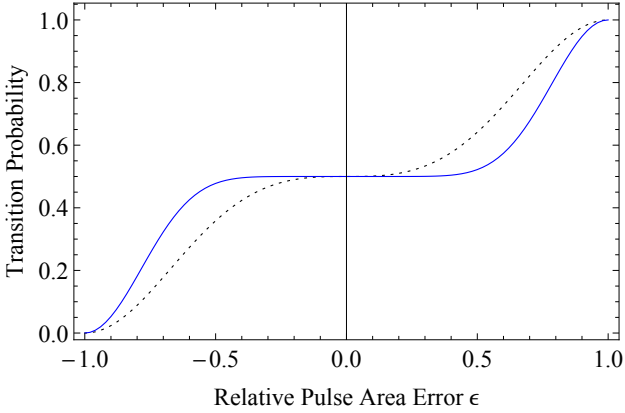}
	\caption{
Comparison of our asymmetric sequence \eqref{half-asym-5} (solid blue), Wimperis' BB1 sequence \eqref{WimpCP} (dotted), and our sequence \eqref{phases-3d} (solid black).
}
	\label{Fig:WimpersiBB1}
\end{figure}

In Fig.~\ref{Fig:WimpersiBB1} we compare the transition profiles of Wimperis' BB1 composite pulse with our sequence \eqref{half-asym-5}, as they both have a total pulse area of $4.5\pi$.
As seen from the figure, our sequence clearly outperforms BB1.
Moreover, even our three-pulse sequence \eqref{phases-3d} outperforms BB1, although it has a total pulse area of just $2\pi$.
However, we should note that the BB1 sequence is a constant-rotation CP.
Hence, as mentioned in the Introduction, it compensates not only pulse-area errors in the transition probability, but also in the phases of the created superposition state.
As such, it is more demanding and requires a longer sequence for the same order of compensation.
On the contrary, our composite sequences produce variable rotations, which compensate pulse area errors in the transition probability only.

\subsection{Nested $\theta$ pulses}

Levitt and Ernst have constructed a family of concatenated CPs by a recursive procedure \cite{Levitt1983}, starting from the two-pulse sequence \eqref{Freeman1980}.
Their composite pulses produce the same profiles as our $\theta$ pulses, when sequences with the same total area are compared.
For instance, there four-pulse $\pi/2$ sequence
\be\label{Levitt-Ernst-4}
A_0 A_{-\frac12\pi} A_0 A_{\frac12\pi}
\ee
is identical to our \eqref{phases-4a-half} sequence.
Their eight-pulse $\pi/2$ composite sequence,
\be
A_0 A_{\frac32\pi} A_0 A_{\frac12\pi} A_\pi A_{\frac12\pi} A_0 A_{\frac12\pi},
\ee
can be obtained from Eq.~\eqref{Levitt-Ernst-4} by our twinning procedure described above, etc.
However, the nesting procedure described in \cite{Levitt1983} is less flexible than our approach as it allows only for total nominal pulse areas equal to $2^n\pi$, with $n=1,2,3,\ldots$.
For comparison, our approach allows one to construct composite $\theta$ pulses with arbitrary total nominal pulse area $N\pi$ with $N=1,2,3,\ldots$.


\section{Comments and conclusions\label{Sec:conclusion}}

We presented a number of composite pulse sequences, which produce arbitrary pre-defined rotations of a two-state system on the Bloch sphere, which are robust to experimental errors in the pulse amplitude and duration.
We derived two main classes of composite sequences: prime and twin sequences.
The prime sequences were derived directly from the propagator, analytically for sequences of two to four pulses and numerically for more than four pulses.
A special attention was devoted to $\pi/2$ sequences, for which general formulas for the phases were derived in two cases of symmetric and asymmetric sequences composed of arbitrarily many pulses.
This allows arbitrarily accurate $\pi/2$ composite pulses to be constructed.
These $\pi/2$ composite sequences were then used to construct three classes of arbitrarily accurate composite $\theta$ pulses by pairing two $\pi/2$ composite sequences, one of which is shifted by phase $\pi-\theta$ with respect to the other one.

In all cases the transition probability of each composite sequence is given by an extremely simple formula which allows one to estimate the accuracy of error compensation.
The order of error compensation is proportional to the nominal total pulse area of the respective sequence.
Comparison of our composite sequences with earlier composite $\theta$ pulses shows equal or better performance.

The results in this paper will be interesting for studies in which only the value of the transition probability (or the population inversion $w$) is of interest, while the behavior of the coherences $u$ and $v$ is of little significance.
Such interest is found in various situations.
For example, when measuring the dependence of the collision rate on the atomic density in a trapped ultracold atomic ensemble one wishes to release a controlled and well-defined fraction of the atoms from the trap \cite{Gerginov}.
The $\theta$ pulses presented here serve this purpose because they can transfer a fraction of the atoms from the trapping state to a non-trapping state with high precision.
Likewise, one can use the composite $\theta$ pulses as a highly accurate output coupler for a pulsed atom laser derived from a Bose-Einstein-condensate \cite{Mewes1997,Andrews1997}.
Another example is found in Ramsey spectroscopy, wherein the precision can be enhanced if the two $\pi/2$ pulses are replaced by two composite $\pi/2$ pulses.
If the latter pulses are identical then the (non-robust) phase of the created superposition is unimportant as it cancels out.
%

The results presented in this work demonstrate the remarkable flexibility of composite pulses accompanied by extreme accuracy and robustness to errors --- three features that cannot be achieved together by any other coherent control technique.


\acknowledgments
This work is supported by the Bulgarian Science Fund Grant No. DO02/3 (ERyQSenS).



\begin{thebibliography}{99}
	
	
	\bibitem{Levitt1979}
	M. H. Levitt and R. Freeman, J. Magn. Reson. \textbf{33}, 473 (1979); 
	
	\bibitem{Freeman1980}
	R. Freeman, S. P. Kempsell, and M. H. Levitt, J. Magn. Reson. \textbf{38}, 453 (1980).
	
	\bibitem{Levitt1982}
	M. H. Levitt, J. Magn. Reson. \textbf{48}, 234 (1982).
	
	\bibitem{Levitt1983} M. H. Levitt and R. R. Ernst, J. Magn. Res. \textbf{55}, 247 (1983).
	
	\bibitem{Tycko1984} R. Tycko and A. Pines, Chem. Phys. Lett. \textbf{111}, 462 (1984).
	
	\bibitem{Tycko1985} R. Tycko, A. Pines and J. Guckenheimer, J. Chem. Phys. \textbf{83}, 2775 (1985).
	
	\bibitem{Shaka1984} A. J. Shaka and R. Freeman, J. Magn. Reson. \textbf{59}, 169 (1984).
	
	\bibitem{Levitt1986} M. H. Levitt, Prog. NMR Spectrosc. \textbf{18}, 61 (1986).
	
	\bibitem{Wimperis1994} S. Wimperis, J. Magn. Reson. \textbf{86}, 46 (1990); \ibid \textbf{109}, 221 (1994).
	
	\bibitem{ions}
	S. Gulde, M. Riebe, G. P. T. Lancaster, C. Becher, J. Eschner, H. H\"{a}ffner, F. Schmidt-Kaler, I. L. Chuang and R. Blatt, Nature \textbf{421}, 48 (2003); 
	F. Schmidt-Kaler, H. H\"{a}ffner, M. Riebe, S. Gulde, G. P. T. Lancaster, T. Deuschle, C. Becher, C. F. Roos, J. Eschner, and R. Blatt, Nature \textbf{422}, 408 (2003); 
	H. H\"{a}ffner, C. F. Roos, R. Blatt, Phys. Rep. \textbf{469}, 155 (2008);
	N. Timoney, V. Elman, S. Glaser, C. Weiss, M. Johanning, W. Neuhauser, and C. Wunderlich, Phys. Rev. A \textbf{77}, 052334 (2008);
	T. Monz, K. Kim, W. H\"{a}nsel, M. Riebe, A. S. Villar, P. Schindler, M. Chwalla, M. Hennrich, and R. Blatt, Phys. Rev. Lett. \textbf{102}, 040501 (2009).
	
	\bibitem{Ivanov2011} S. S. Ivanov and N. V. Vitanov, Opt. Lett. \textbf{36}, 7 (2011).
	
	\bibitem{Ivanov2015} S. S. Ivanov and N. V. Vitanov, Phys. Rev. A \textbf{92}, 022333 (2015).
	
	\bibitem{Torosov2011PRA} B. T.~Torosov and N. V.~Vitanov, Phys. Rev. A \textbf{83}, 053420(7) (2011).
	
	\bibitem{Torosov2011PRL} B. T.~Torosov, S. Gu\'erin and N.V.~Vitanov, Phys. Rev. Lett. \textbf{106}, 233001 (2011).
	
	\bibitem{Schraft2013} D. Schraft, T. Halfmann, G. T. Genov, and N. V. Vitanov, Phys. Rev. A \textbf{88}, 063406 (2013).
	
	\bibitem{Genov2014PRL}
	G. T. Genov, D. Schraft, T. Halfmann and N. V. Vitanov, Phys. Rev. Lett. \textbf{113}, 043001 (2014).
	
	\bibitem{West1949} C. D. West and A. S. Makas, J. Opt. Soc. Am. \textbf{39}, 791 (1949).
	
	\bibitem{Destriau1949} M. G. Destriau and J. Prouteau, J. Phys. Radium \textbf{10}, 53 (1949).
	
	\bibitem{Pancharatnam1955} S. Pancharatnam, Proc. Ind. Acad. Sci. \textbf{51}, 130 (1955); ibid. \textbf{51}, 137 (1955).
	
	\bibitem{Harris1964} S. E. Harris, E. O. Ammann, and A. C. Chang, J. Opt. Soc. Am \textbf{54}, 1267 (1964).
	
	\bibitem{McIntyre1968} C. M. McIntyre and S. E. Harris, J. Opt. Soc. Am \textbf{58}, 1575 (1968).
	
	\bibitem{Peters2012}
	T. Peters, S. S. Ivanov, D. Englisch, A. A. Rangelov, N. V. Vitanov, and T. Halfmann,
	Appl. Opt. \textbf{51}, 7466 (2012).
	
	\bibitem{Ivanov2012}
	S. S. Ivanov, A. A. Rangelov, N. V. Vitanov, T. Peters, and T. Halfmann,
	J. Opt. Soc. Am. A  \textbf{29}, 265 (2012).
	
	\bibitem{Kyoseva2013} E. Kyoseva and N. V. Vitanov, Phys. Rev. A \textbf{88}, 063410 (2013).
	
	\bibitem{Vitanov2011} N. V. Vitanov, Phys. Rev. A \textbf{84}, 065404 (2011).
	
	\bibitem{Torosov2018} B. T.~Torosov and N.V.~Vitanov, Phys. Rev. A \textbf{97}, 043408 (2018).
	
	\bibitem{Gerginov} V. Gerginov, private communication.
	
	\bibitem{Mewes1997} 
	M.-O. Mewes, M. R. Andrews, D. M. Kurn, D. S. Durfee, C. G. Townsend, and W. Ketterle, Phys. Rev. Lett. \textbf{78}, 582 (1997).
	
	\bibitem{Andrews1997}
	M. R. Andrews, C. G. Townsend, H.-J. Miesner, D. S. Durfee, D. M. Kurn, W. Ketterle, Science \textbf{275}, 637 (1997).
	
	
	
\end{thebibliography}
\end{document}